\documentclass[autheryear, times]{elsarticle}

\usepackage{geometry}
\usepackage[]{algorithm2e}
\usepackage{self}
\usepackage{amsthm}
\usepackage{times}
\usepackage{url}
\usepackage{natbib}
\usepackage{amsmath}
\usepackage{longtable}
\usepackage{amsfonts}
\usepackage{amssymb}
\usepackage{graphicx}
\usepackage{multirow}
\usepackage{subcaption}
\usepackage{verbatim}
\usepackage{lineno}
\usepackage{natbib}
\usepackage[ngerman]{varioref}
\usepackage{hyperref}
\usepackage{cleveref}
\biboptions{authoryear}
\usepackage{subcaption}
\usepackage{multirow}
\usepackage{tikz}
\usetikzlibrary{shapes.geometric, arrows}
\tikzstyle{startstop} = [rectangle, rounded corners, minimum width=3cm, minimum height=1cm,text centered, draw=black, fill=red!30]
\tikzstyle{io} = [trapezium, trapezium left angle=70, trapezium right angle=110, minimum width=3cm, minimum height=1cm, text centered, draw=black, fill=blue!30]
\tikzstyle{process} = [rectangle, minimum width=3cm, minimum height=1cm, text centered, draw=black, fill=orange!30]
\tikzstyle{decision} = [diamond, minimum width=3cm, minimum height=1cm, text centered, draw=black, fill=green!30]
\tikzstyle{arrow} = [thick,->,>=stealth]

\graphicspath{{img/}}

\journal{Transportation Research Part C}

\begin{document}

\begin{frontmatter}
	
	\title{User-centric interdependent urban systems: using time-of-day electricity usage data to predict morning roadway congestion}

	\author[mymainaddress]{Pinchao Zhang}
	\ead{pinchaoz@andrew.cmu.edu}
	\author[mymainaddress,mymainaddress2]{Zhen (Sean) Qian\corref{mycorrespondingauthor}}
	\cortext[mycorrespondingauthor]{Corresponding author}
	\ead{seanqian@cmu.edu}

	\address[mymainaddress]{Department of Civil and Environmental Engineering, Carnegie Mellon University, Pittsburgh, PA 15213}
		\address[mymainaddress2]{Heinz College, Carnegie Mellon University, Pittsburgh, PA 15213}
	
	\begin{abstract}
Urban systems are interdependent as individuals' daily activities engage using those urban systems at certain time of day and locations. There may exist clear spatial and temporal correlations among usage patterns across all urban systems. This paper explores such a correlation among energy usage and roadway congestion. We propose a general framework to predict congestion starting time and congestion duration in the morning using the time-of-day electricity use data from anonymous households with no personally identifiable information. We show that using time-of-day electricity data from midnight to early morning from 322 households in the City of Austin, can make reliable prediction of congestion starting time of several highway segments, at the time as early as 2am. This predictor significantly outperforms a time-series predictor that uses only real-time travel time data up to 6am. We found that 8 out of the 10 typical electricity use patterns have statistically significant affects on morning congestion on highways in Austin. Some patterns have negative effects, represented by an early spike of electricity use followed by a drastic drop that could imply early departure from home. Others have positive effects, represented by a late night spike of electricity use possible implying late night activities that can lead to late morning departure from home.
	\end{abstract}
	
	
\end{frontmatter}
\modulolinenumbers[5]

\section{Introduction}
\label{sec:1}
\begin{figure}[h]
\centering
\includegraphics[scale=0.7]{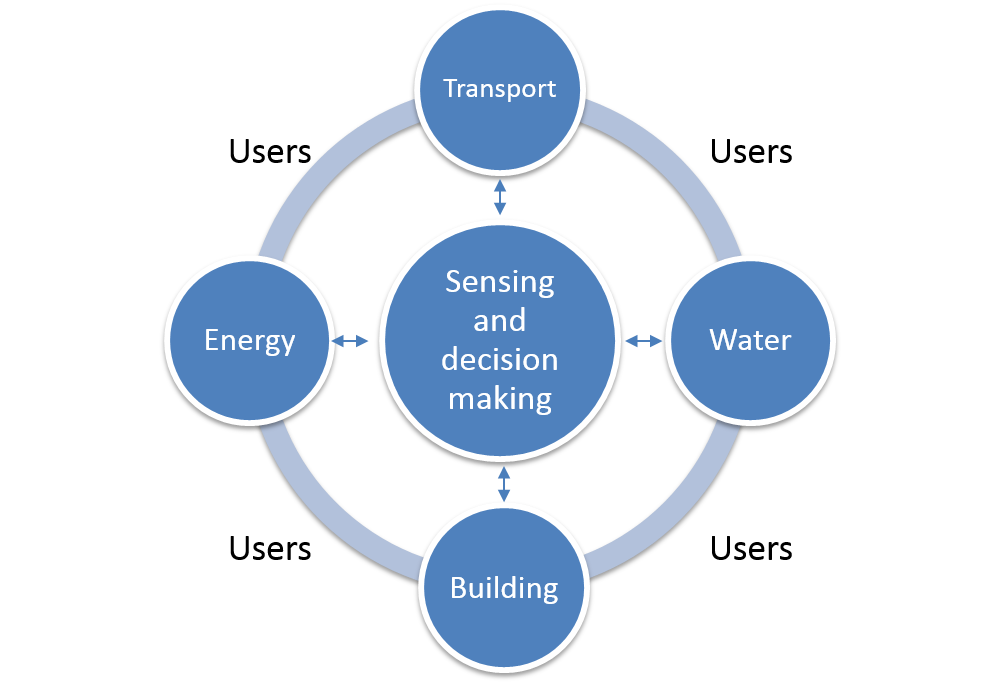}
\caption{ Interdependency of some urban systems: their system user patterns are inter-related both temporally and spatially}
\label{Fig:inter}
\end{figure}

Central to smart cities is the complex nature of interrelationships among various urban systems. Linking all urban systems is the system users. The individual daily activities engage using those urban systems at certain time of day and locations. There may exist clear spatial and temporal correlations among usage patterns across all urban systems. A general idea is to fuse and analyze user demand data from transportation, energy, water and building systems (as shown in Figure \ref{Fig:inter}) to discover the spatio-temporal usage patterns among those systems. This enables cross-system demand prediction and management. For some users, the usage of one urban system is likely to be used minutes or hours ahead of their usage of other urban system(s) as a result of daily activity chains. Therefore, the spatio-temporal usage of a urban system can be accurately predicted a few minutes or hours ahead by real-time sensing user patterns of other urban system(s). This is otherwise hard to accomplish by solely monitoring one ``siloed'' system. Ultimately, real-time control strategies for demand management of one urban system can be developed with efficient real-time demand prediction upon other urban system(s).

Following this general idea, the objective of this paper is to explore spatio-temporal correlations of usage patterns between energy systems and transportation systems. Specifically, as a first attempt, we propose a methodology along with data analytics for the City of Austin to address the following two questions. What can we tell about the morning commute by knowing households’ electricity use the night before or early in the morning? How would the households' electricity use data help predict morning congestion in the real-time (or a few hours early) comparing to using real-time traffic data only?

\begin{figure}[h]
        \begin{subfigure}[b]{0.49\textwidth}
        \includegraphics[width=\textwidth]{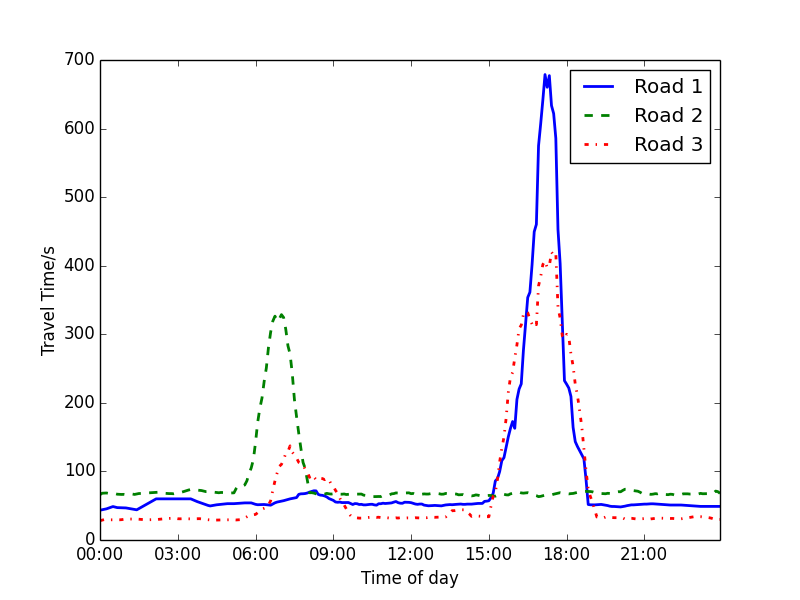}
        \caption{}
        \label{ss trend}
    \end{subfigure}
            \begin{subfigure}[b]{0.49\textwidth}
        \includegraphics[width=\textwidth]{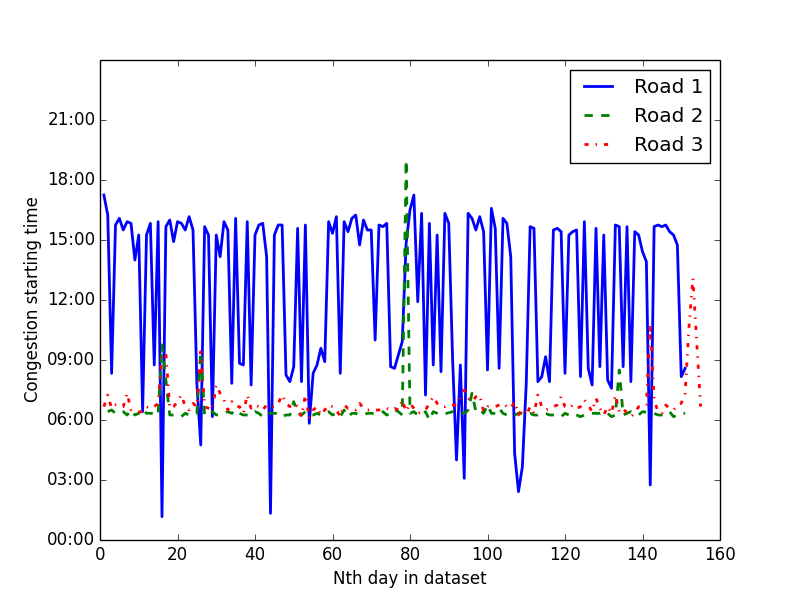}
        \caption{}
        \label{ss portion}
    \end{subfigure}
\caption{On weekdays, congestion varies from day to day: (a) Time-varying travel times of three road segments on a same day; (b) Congestion starting time of three road segments}
\label{fig:eg}
\end{figure}

To conceptually demonstrate why real-time traffic data is usually not sufficient for real-time traffic prediction, we obtain travel speed data for three typical highway segments on I-35 in the City of Austin. On a typical day, congestion occurs in the afternoon peak for Segment 1, but generally not in the morning. Segment 2 typically has morning peak congestion, but not in the afternoon. Segment 3 has both morning and afternoon congestion in most of days. Figure \ref{fig:eg}(a) plots their respective time-varying travel times (in seconds) on a typical weekday (Jan 08, 2014). The free-flow traffic/passenger flow in the early morning does not exhibit clear patterns before it transitions to being congested (also known as traffic “break-down”). For all the three segments, the travel time stays flat (namely in free flow) until traffic break-down that causes an instantaneous drop in speed. Real-time monitoring the speed or travel time does not necessarily help predict the exact time of traffic break-down, nor would historical data help as much due to day-to-day variation. If we define ``congestion starting time'' as the time when traffic speed reduces by 50\% over 10 minutes, Figure \ref{fig:eg}(b) shows the congestion starting time of those three road segments for 155 weekdays in 2014. For segment 3, the morning congestion starting time varies by 30-60 mins from day to day. The day-to-day variation of morning congestion patterns on both segments 1 and 2 are less than segment 1. Morning congestion occurs for about 20\% of days on segment 1. Congestion started with 6:10-6:20am for most days on segment 2, but there are nine days when its congestion started after 6:30am.  To sum up, daily congestion patterns are difficult to predict by only monitoring real-time traffic flow because traffic break-down is very sensitive to supply/demand that is usually random on the daily basis. Historical information also does not seem to help much due to substantial day-to-day variation.

When taking energy systems into consideration, there may exist spatio-temporal relations between the travel demand and electricity demand. When the congestion starts and how long it lasts is a result of demand characteristics on each day, namely what time travelers leave home and their travel/driving behavior. Those daily characteristics may be partially attributed to those commuters' activities at midnight or early in the morning (such as 3-5am). Therefore, real-time traffic prediction can be complemented and enhanced by mining additional real-time electricity usage data in addition to traffic data. We hope to provide more accurate prediction for real-time traffic. In principle, we would expect electricity usage data to add additional insights for a better traffic prediction, due to spatially and temporally correlated user activities on both roadway and energy systems.


This paper presents a generic modeling framework to mine time-varying usage data of one system to predict usage of another system. Specifically, it addresses real-time traffic prediction using midnight and early morning electricity usage data. Conceptually, the prediction algorithm first discovers users categorized by their time-of-day electricity use patterns, learns the most critical features from those user categories that are related to traffic congestion, and finally builds a predictor dependent on electricity use features with parameters calibrated by real data. Our results show that, for some roadway segments, using electricity usage data can provide more robust and accurate prediction of when congestion starts in the morning, comparing to the prediction using traffic data only.

The rest of this paper is organized as follows.  In \cref{sec:2}, related work and applications on energy use and short-term traffic prediction are reviewed. \Cref{sec:3} discusses the methodology proposed in this paper. The dataset and data processing are introduced in \cref{sec:4}. Experiment results are shown and discussed in \cref{sec:5}. Finally \Cref{sec:6} concludes this paper.

\section{Literature review}
\label{sec:2}

Electricity consumption by individuals is related to many factors, such as usage purposes, weather, and socio-demographics of users. Decades ago, collection of household-level energy use data was challenging and required substantial man power. With the introduction of automated metering infrastructure (AMI), electricity use meters are integrated and connected. Meter data started to be managed, archived, accessed, shared and better used for decision making of smart grids.  \cite{rhodes2014experimental} explained how grid data was collected through smart grid meter systems in Austin, TX across electricity, gas and water systems.  Having access to this data has direct societal benefits to save energy. For example,  \cite{Schultz2015} gave real-time feedback to users on the electricity consumption and help users save energy use based on data collected by smart meters.  Their results showed that simply displaying current consumption and current average consumption of similar households could help users reduce electricity use over 10\%.

AMI data have been intensively utilized to learn demand features for better understanding, predicting, and incentivizing demand. For instance, \cite{barker2014empirical} captured key characteristics of each load type (e.g. microwave, laundry machines) from the data and proposed a method to retrieve load types by time of day. \cite{dyson2014using}  used the smart meter data together with temperature data to predict residential air conditioning use. \cite{espinoza2005short} proposed a Periodic Autoregression model in a unified time-series framework for short-time forecast of electricity use.  \cite{tso2007predicting} conducted a comparative study of electricity use prediction using regression analysis, decision trees, and neutral networks respectively, and found that no single method is always better than others. In fact, they showed that season largely determines energy use. It seems natural to classify users into different groups, each of which implies a typical electricity use pattern \cite[e.g.][]{chicco2006comparisons,flath2012cluster, habenanalysis,chicco2006comparisons}.  The general idea is to group users with similar time-of-day energy use patterns. The time-of-day energy use is also known as a daily profile of a user.  Grouping users by patterns provide insights for policymakers and utility companies to efficiently incentive users to save energy and improving social welfare. \cite{mohsenian2010autonomous} focused on demand-side management and proposed an incentive-based energy consumption scheduling algorithm to minimize total energy use while balancing the load over time.

Note that those utility consumers are most likely to travel. Their electricity use patterns in aggregation may be partially related to travel demand characteristics, namely when, how and where to travel.  Those studies regarding energy demand response show promise to understand cross-system users' characteristics that can be potentially utilized to predict transportation system use.


Predicting short-term traffic conditions is of long-time interest in transportation research literature since it plays a pivotal role in traffic management and operations. They have been extensively explored. The predicted metrics representing traffic conditions and performance include traffic flow/counts \citep[e.g.,][]{ghosh2009multivariate,ghosh2007bayesian}, density \citep{anand2014data,padiath2009prediction}, and speed/travel time \citep{oda1990algorithm,ishak2002performance}.

The prediction methodologies can be categorized into two types. The first type is based on theoretical models and simulation of traffic flow, such as a pheromone model \citep{ando2005pheromone}, the Cell Transmission Model in conjunction with a seasonal autoregressive integrated moving average model \citep{szeto2009multivariate}, a Kalman filter embedded with a first-order LWR traffic flow \citep{work2010traffic}, and a Dynamic Stochastic Assignment Model \citep{ben1992real}. The traffic dynamics was represented by traffic flow models which are corrected as time progresses by observations. This type of traffic prediction has a strong theory support. However, it generally does not work for large-scale networks constrained by computational complexity of real-time flow simulation. In addition, the prediction results may be far off when the theoretical flow dynamics is not bounded by data that are very sensitive to any disruption to demand and supply.

The second type of short-term traffic prediction relies solely on statistical or machine learning models applied to time-varying traffic speed data, such as time series models \citep{ghosh2009multivariate,ghosh2007bayesian}, graphical models \citep{sun2006bayesian,sun2004short} and deep learning/neural networks \citep{kumar2013short,lv2015traffic}. The prediction model needs to be well trained or calibrated before it makes real-time prediction. Generally, the prediction performance is fair when no substantial network disruption occurs or under network disruptions that are intensively learned from training data. This type of model cannot deal with network flow anomaly, nor under ``what-if'' scenarios. This is simply because it cannot fully adapt to traffic conditions in large-scale networks that are little covered by observations.

Distinct from previous studies in the literature, this paper uses time-of-day electricity data at the household level at midnight or early in the morning to predict morning-peak traffic. To our best knowledge, this is the first study that attempts to discover spatio-temporal relations of usage patterns among transportation and energy systems. The relations can infer users' behavior and are further utilized to predict traffic in the real-time. Instead of working directly with prediction of traffic speed, density or flow rate, the target of this paper is to predict if there will be congestion occurring in the morning peak, and if so, what exact time will the congestion start and how long it will last. This is a critical measure for both travelers and system managers. As we can see from Figure \ref{fig:eg} that some road segments may have morning congestion occasionally, which can be hard to predict ahead of time. Other segments have morning congestion on every weekday, but the congestion starting time and duration vary quite a bit from day to day. Our hope is that the day-to-day variation can be partially explained by users' daily activities reflected from electricity use.


\section{Methodology}
\label{sec:3}
In this section, we discuss a methodology to predict usage of one system from mining data from another system in real time. The methodology can be generically applied to any interdependent systems. It predicts the usage of System A at time $t$ from the usage patterns of System B spanning a time-of-day period from $t'$ to $t''$, where the time interval $[t',t'']$ is prior to $t$ on the same day. Some historical information is also learned from data and incorporated into the predictor. The predictor has to be well trained by data of both systems observed over a fair number of days. To use this predictor, at time $t''$ on each day, the usage of System B would have been observed from $t'$ to $t''$. Those observations can be input to the predictor, at time $t''$, to forecast the usage of System A for the future time $t$.

In this paper, we specifically consider predicting traffic congestion at a morning time $t$, using household-level electricity usage data during the time interval $[t',t'']$ that can be up to a few hours earlier than $t$ on the same day. As we will show later, this predictor can outperform a predictor using only the traffic data, even at the time much closer to $t$ than $t''$. This is because morning traffic break-down is largely attributed to random demand characteristics, such as demand level, driving behavior, departure time from home, etc. Those demand characteristics can hardly be predicted using traffic data only in the real time, and oftentimes the traffic break-down may occur in the morning peak without sufficient prior signs from traffic data. Our hope is that electricity use at midnight or in the very early morning can partially reveal those demand characteristics to some extent, and therefore can help better predict traffic congestion for some locations.

Consider electricity use data at the household level. On each day, $e^d_h(t)$ denotes the electricity use at time of day $t$ on day $d$ for the $h$-th household. Suppose we collected data for $H$ households, $D$ days, and $T$ time-of-day intervals. $e^d_h\doteq \{e^d_h(t)\}_t$ is a $T\times 1$ vector, also known as a daily \textbf{profile} of electricity use on day $d$ for the $h$-th household. There are in all $DH$ daily profiles. Similarly, $e^d\doteq \{c^d_h\}_h$ is a $TH\times 1$ vector, namely daily observations for electricity use. We intend to predict both ``congestion starting time'' and ``congestion duration'' on a road segment, denoted by $k^d$ and $c^d$ on day $d$, where both $k^d$ and $c^d$ are at a much later time than those $T$ time intervals of electricity use observations. Our goal is to establish,
\begin{equation}\label{pred}
  k^d = f(e^d;\beta), c^d = f'(e^d;\beta')
\end{equation}
Where $f(\cdot)$ and $f'(\cdot)$ is a model mapping electricity use to congestion starting time and duration, respectively, and $\beta,\beta'$ are a vector of parameters to be learned from data.

The intuition behind $f(\cdot)$ and $f'(\cdot)$ is that electricity use from midnight to 5:30am may reveal demand characteristics, namely what time, and how many travelers depart home for work. For instance, electricity use of a household drops when family members go to sleep, or it drop when they depart home. However, due to complicated spatio-temporal relationship among electricity use and transportation system use, $f(\cdot), f'(\cdot)$ are generally unknown and difficult to construct using real-world physics. Therefore, we use data mining models (i.e., regression models) to estimate $\theta,\theta'$ for assumed functions $f(\cdot)$ and $f'(\cdot)$.

There are many different ways to construct functions $f(\cdot)$ and $f'(\cdot)$. Complicated data mining models such as Neural Networks can deal with high dimensional data, but can be notoriously difficult to interpret. In order to obtain insights regarding cross-system interrelations, in this paper, we use simple linear predictors in conjunction with an unsupervised learning approach.



\subsection{Clustering}

We suppose there exist some typical profiles of the electricity use, namely \textbf{daily patterns}, that characterize individuals and their daily activities. It is expected the influence of each pattern of individuals and their daily profiles to the traffic congestion is stable from day to day, and can be mined using historical data.

Some intuitive examples include,
\begin{itemize}
  \item One daily pattern could be that there is a spike of electricity use from midnight to 2am, while the use is low during the morning til 6am. This implies that the group of users under this pattern may not commute during the morning peak, and therefore they tend to not contribute to the congestion.
  \item A second daily pattern could be that electricity use of a user is low from midnight to 5:30am, followed by a rise after 5:30am. This is perhaps a typical ``morning commute'' pattern for a typical commuter.
  \item Another daily pattern may be that the electricity use steadily declines over time from midnight to 5:30am, as a result of air conditioning running all night during the summer. Then the use rises at 5:30am for commuters' morning activities.
\end{itemize}

Some daily patterns are straightforward to relate to users' activities. However, electricity use can be attributed to many other factors that are random, noisy, and hard to interpret. Therefore, we use a data-driven approach to separate observations into ``clusters'' or ``typical patterns'' that are representative of typical electricity use among all sampled users. The increase and decline of electricity use over time for all patterns will be determined by the statistical models rather than manual specifications.

Another note to make here is that a user can switch among those typical patterns from day to day. A user may be a commuter, namely under a typical commuter daily pattern, from Mondays to Thursdays, while behaving under a non-commuting daily pattern on Fridays. She does not contribute to the morning-peak congestion on Fridays. We would like to use data mining models to capture the characteristics of individuals and their patterns on each day, both of which are used to predict morning congestion.

In this paper we use the simple k-means clustering algorithm to learn daily patterns. We have $H$ households and their daily profiles of $D$ days, in all $DH$ samples: $\{e_h^d\}_{d,h}$. We partition the samples into $K$ clusters (or typical patterns) denoted as pattern $1$, pattern $2$, ... , pattern $K$. Each pattern $i$ is represented by the centroid of its cluster (usually the numerical average) $e_i$. The number of clusters $K$ must be predefined before performing the clustering. The GAP statistics (\cite{tibshirani2001estimating}) can be used to determine the value of $K$.



\subsection{A linear predictor}

We assume that congestion starting time on the $d$-th day is predicted by its expected value that is a linear combination of many features,
\begin{align}
\mathbb{E}(k^d) = \beta^T  x^d 
\end{align}
Where $x^d$ is a vector of $p$ features observed on the $d$-th day. $x^d = [x_1^d,x_2^d,...,x_p^d]$. Those features are derived from daily profiles $e^d$. The commonly used ordinary least square (OLS) linear predictor is to learn the parameters $\beta$ such that

\begin{align}
\min_{\beta} ||k-x^{\top}\beta||_2
\end{align}


However, if the dimension of $x$ becomes very large or larger than the sample size, then the regression tends to overfit the data. To overcome this issue, we use both L1-norm penalization and cross validation. The first idea is to add a L1-norm regularization of $\beta$ (LASSO, \cite{tibshirani1996regression} ) and the original problem becomes,
\begin{align}
\min_{\beta} ||k-x^{\top}\beta||_2+\alpha ||\beta||_1\label{lasso}
\end{align}
$||\beta||_1 = \sum_i |\beta_i|$ denotes the $L1$ norm of $\beta$. The LASSO regression helps select the most critical features that are linearly related to the response. Second, K-fold cross-validation will be used to measure the performance of our model. 

We construct features on each day that are derived from the clustering results of daily profiles. Note that the k-means algorithms are applied on the $DH$ daily profiles. Thus, each household or user on each day will be assigned to a typical pattern (cluster). Features that are related to traffic congestion can be both aggregate or disaggregate. At an aggregate level, we assume that all households/users under the same typical pattern on each day affect the congestion in the same way (namely all households/users under the same pattern are homogeneous), and speculate that the total number of households/users under each of the typical patterns on each day attributes to congestion. At a disaggregated level, we assume each sampled household/user represents a particular group of users, and each contributes to congestion differently as opposed to the user homogeneity assumption. Therefore, we create features for each sample household/user for the predictor of traffic congestion. In the rest of the paper, we refer to the two types of features as ``aggregate features'' and ``disaggregate features'', respectively.



Aggregate features are highly compressed. They are a vector of $K-1$ elements for each daily profile. Each element is the ratio of households/users that are under a pattern on that day. The element of one last pattern can be dropped from the vector as a result of redundancy.

Those aggregate features would offer effective prediction if the assumption holds that all households/users are homogeneous under the same electricity use pattern. If this is not reasonable, then disaggregate features allow us to examine household/user-level behavior in full details. On each day, the electricity use pattern of a household is represented by a vector consisting of $K-1$ binary elements where an element is 1 if a pattern is followed and zero otherwise. For example, if a household is under the second typical electricity use pattern on a day, then the corresponding vector for this household on this day is $[0,1,\cdots,0]$ (with $K-2$ zeros). After creating the vectors for all households/users, we concatenate all $H$ vectors to form a large disaggregate feature vector for that day. The vector length is $(K-1)\times H$, possibly exceeding the sample size. This is where L1-norm regularization and cross-validation play a role to mitigate overfitting. In principle, we would expect some households/users are more influential to the traffic congestion than others, which can be learned from data.


\section{Data acquisition and pre-processing}
\label{sec:4}
This paper uses two data sets in the City of Austin, Texas collected in 2014. The first is the household-level electricity usage data, and the second is the archived travel time data for highway segments. The ultimate goal is to use electricity use data from midnight to 6 am to predict morning commute traffic congestion on each day. 
\begin{figure}[h]
        \begin{subfigure}[b]{0.49\textwidth}
        \includegraphics[width=\textwidth]{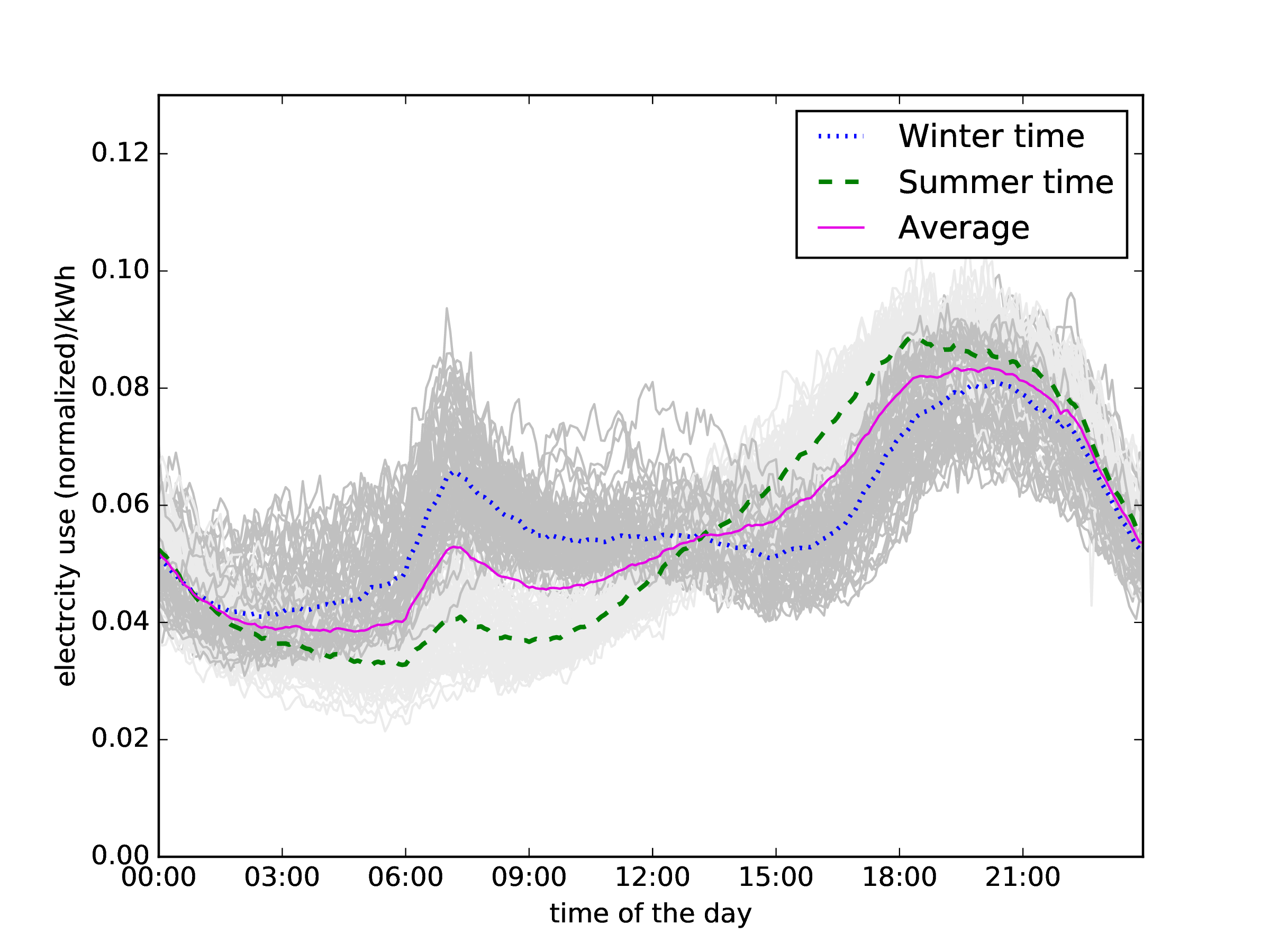}
        \caption{All daily profiles and their averages}
        \label{ss trend}
    \end{subfigure}
            \begin{subfigure}[b]{0.49\textwidth}
        \includegraphics[width=\textwidth]{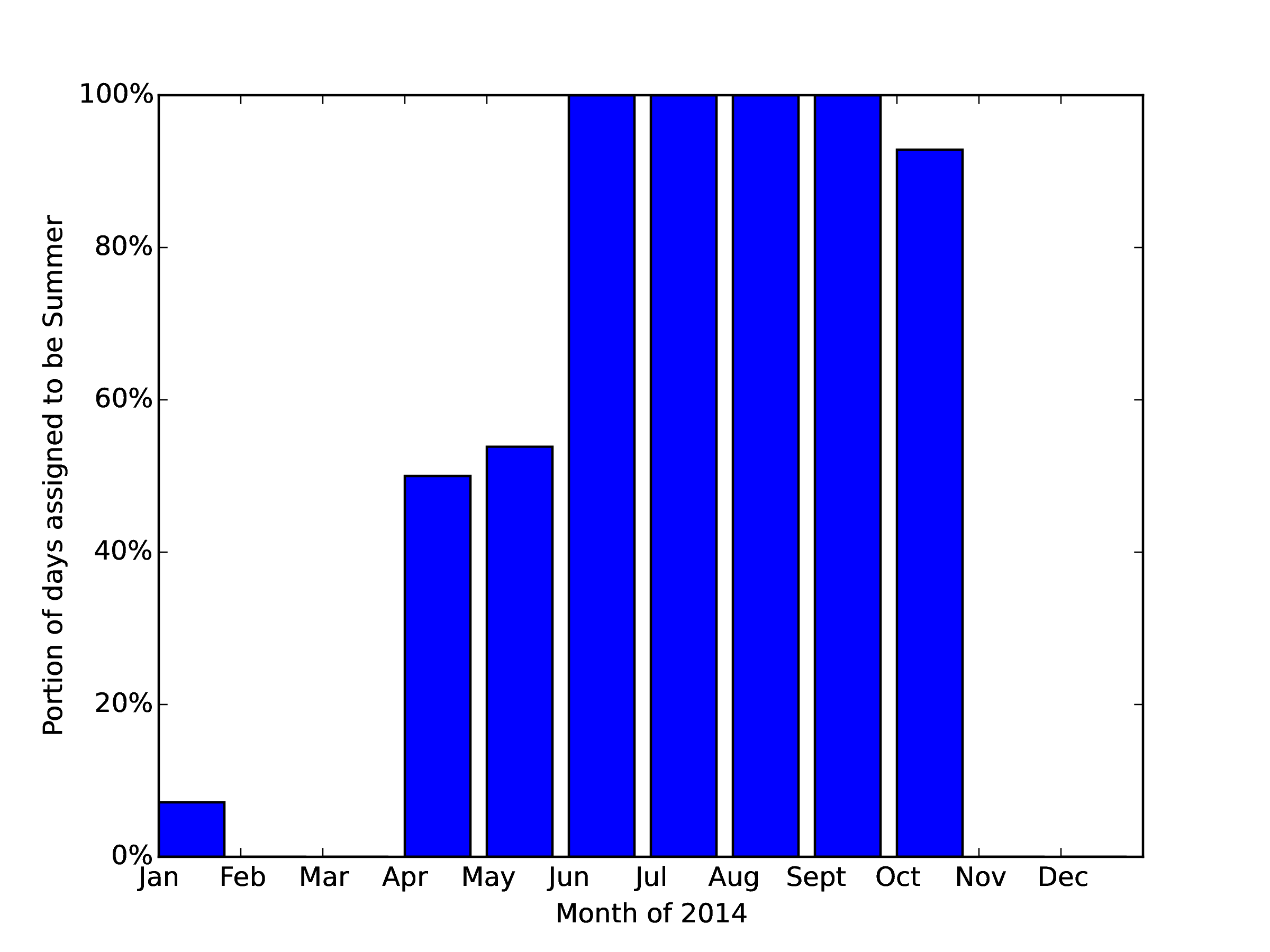}
        \caption{Portion of summer profiles in each month}
        \label{ss portion}
    \end{subfigure}
\caption{(a) All the (normalized)  daily profiles (322 $households$ $\times$ 251 $weekdays$) of the electricity usage are in grey. Light grey are winter profiles and dark grey are summer profiles. The daily profiles were collected from 12:00 am to 11:55 pm on every weekday of 2014. The purple line is the daily average over all weekdays. The green line is the average of all winter weekdays and the blue one is the average of all summer weekdays. (b) shows the portion of daily profiles in each month that are in the summer cluster.}
\label{fig:1}
\end{figure}
\subsection{Electricity usage}

The electricity usage data are acquired from an Advanced Metering Infrastructure (AMI) program run by Pecan Street Inc. There were around 400 households participating in the program in the year of 2014. There were some households entering and leaving this program during the year. We choose those households that stayed the entire year of 2014, in all 322 households. For each of those households, the electricity use (in kWh) were recorded in 5 min time intervals throughout the year. We consider all 251 weekdays in 2014. The daily use profiles from midnight to 6 am are normalized in a way that its sum of squares is one. Thought each household is referred to with a unique reference ID number, those households/users are completely anonymous. Their locations and any other private information are filtered out and unknown to this research.


As suggested by \cite{paatero2006model} and \cite{tso2014multilevel}), the electricity usage often has a seasonal effect. We first use a simple K-means clustering with $K=2$ to separate all workdays into two seasons. \Cref{fig:1} shows the seasonal representatives of the electricity usage, indicating a clear separation of seasonal patterns of summer and winter. A larger portion of electricity is consumed during the night in summer than winter, whereas more electricity is used during the morning in winter than summer. \Cref{ss portion} implies that the clustering can effectively distinguish the two seasons. In this study, we only use weekdays in summer. By clustering, the summer starts in April and ends in October. To completely filter out the seasonal effect, only weekdays from May to October are used. Furthermore, we also found that the daily patterns on Monday and Friday could be very different from those on other weekdays. Hence we focus on all Tuesdays, Wednesdays and Thursdays in summer, in all 79 weekdays.



\begin{figure}[h]
\centering
\includegraphics[scale=0.7]{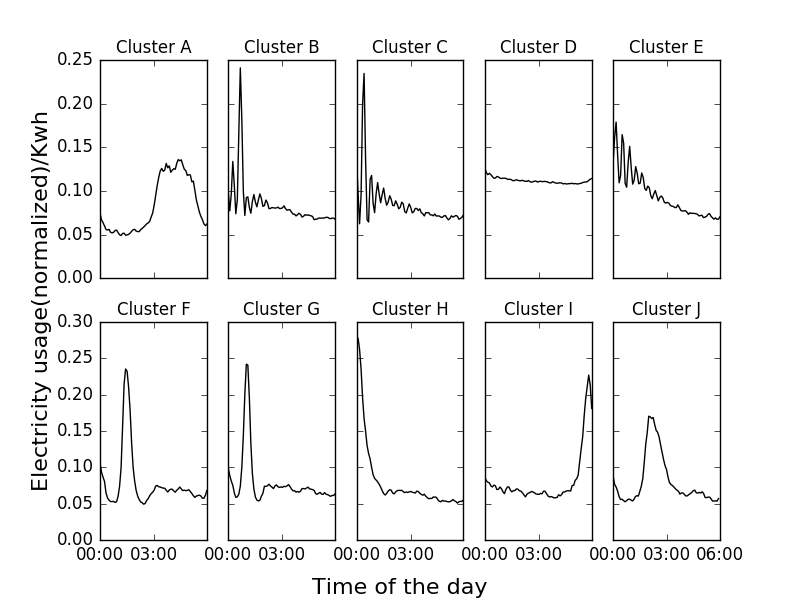}
\caption{The $10$ most representative electricity usage patterns}
\label{c:result}
\end{figure}

Next we conduct a clustering analysis for all $79 \ weekdays\times 322 \  households=25,438$ daily profiles. Each daily profile is a vector of 72 elements representing the electricity use of all 5-min time intervals from 12 am to 6 am.  $K=10 $ is selected by GAP statistics for the K-means algorithm. \Cref{c:result} plots the daily profile of cluster average for each of the $10$ clusters, representing $10$ most representative patterns. We denote the ten clusters as patterns $A$, $B$,$C$,...,$J$. The time-of-day electricity use varies quite substantially among those patterns. For example, pattern $F$ and pattern $G$ both have a peak electricity use for about one hour after midnight. Pattern $E$ implies steadily declining use of electricity after midnight til 6 am, which differs from the dramatic decline to low usage by 1pm in patterns $B, C$ and $H$. Households in pattern $A$ and pattern $I$ are likely to get up early and possibly leave home early for work. Each pattern may imply a specific influence on the traffic congestion, which will be learned from data.

\subsection{Traffic data}
\label{sec:congest}

Historical travel time data are acquired from National Performance Management Research Data Set (NPMRDS).  The travel time data were provided in 5-min time intervals, and cover highways and major roads around the Austin Metropolitan Area. Its spatial resolution is defined by Traffic Management Channels (TMC), and acquired from NPMRDS as well.

\begin{figure}
\centering
    \begin{subfigure}[b]{0.49\textwidth}
        \includegraphics[width=\textwidth]{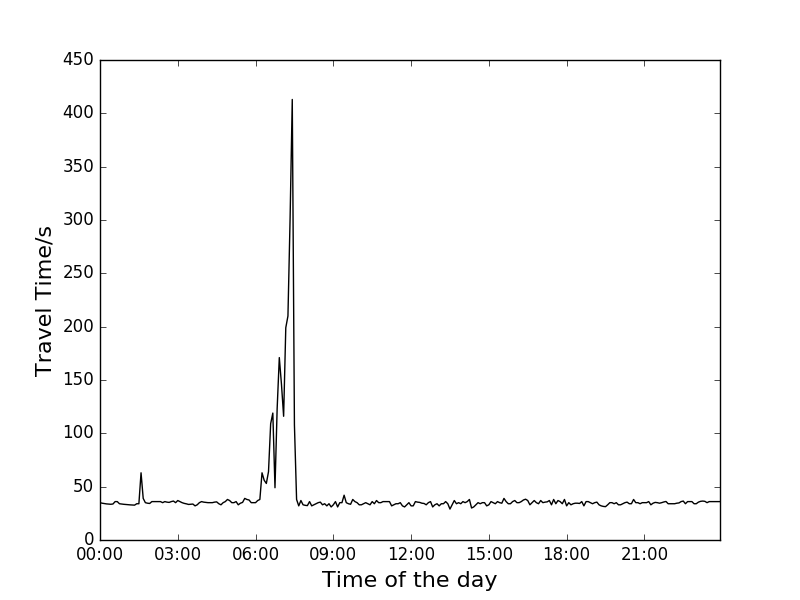}
        \caption{}
        \label{fig:2a}
    \end{subfigure}
    \begin{subfigure}[b]{0.49\textwidth}
        \includegraphics[width=\textwidth]{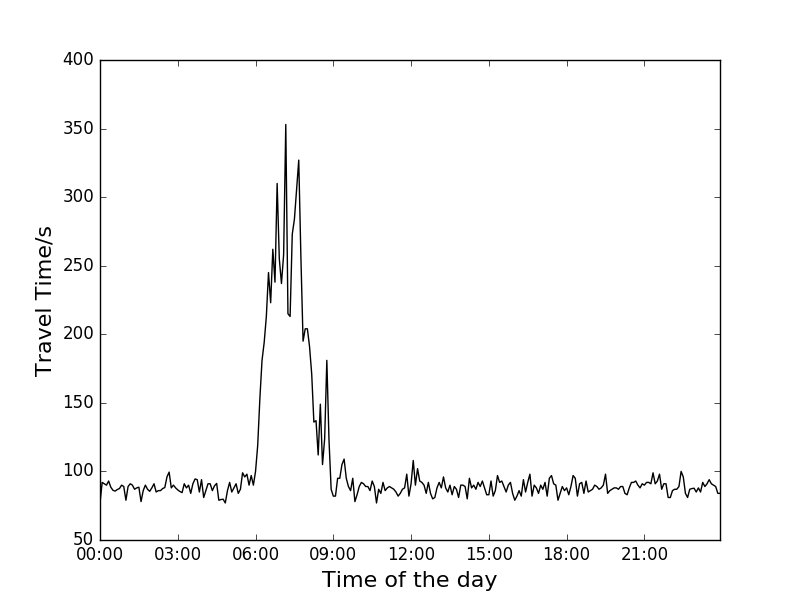}
        \caption{}
        \label{fig:2b}
    \end{subfigure}
    \begin{subfigure}[b]{0.49\textwidth}
        \includegraphics[width=\textwidth]{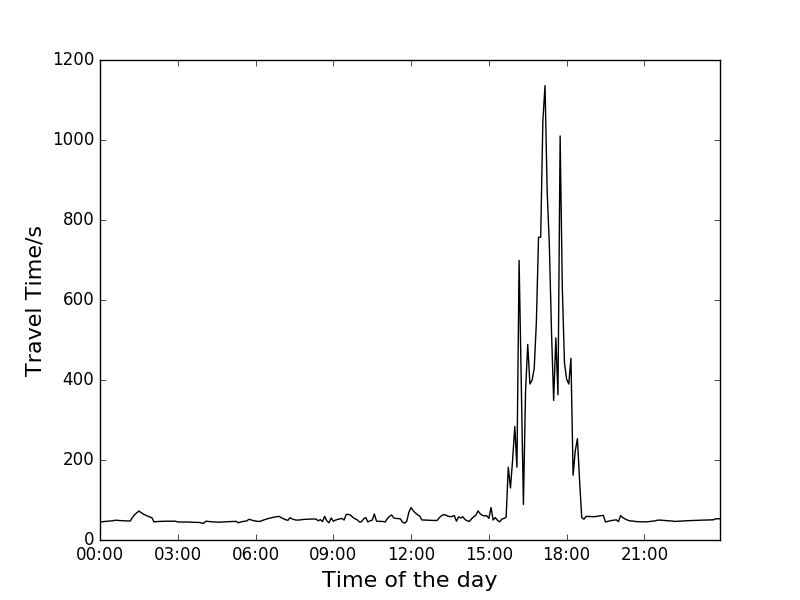}
        \caption{}
        \label{fig:2c}
    \end{subfigure}
        \begin{subfigure}[b]{0.49\textwidth}
        \includegraphics[width=\textwidth]{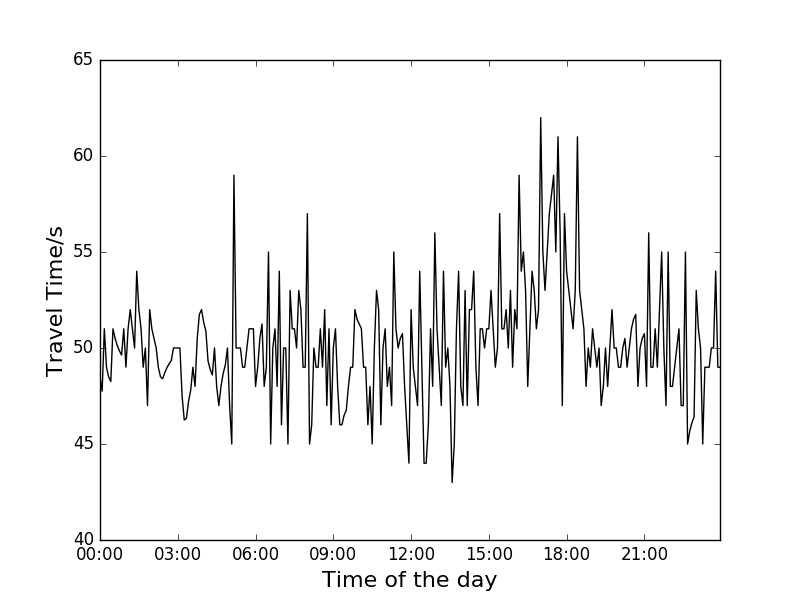}
        \caption{}
        \label{fig:2d}
    \end{subfigure}
\caption{Four travel time patterns: (a) A link with a short morning congestion period on June 30th, 2014; (b) A link with a long morning congestion period on July 1st, 2014; (c) A link with afternoon congestion period on June 30th, 2014; and (d) A link with no clear time-of-day congestion pattern on June 30th, 2014. }
\label{fig:2}
\end{figure}

\begin{figure}
\centering
\includegraphics[scale=0.7]{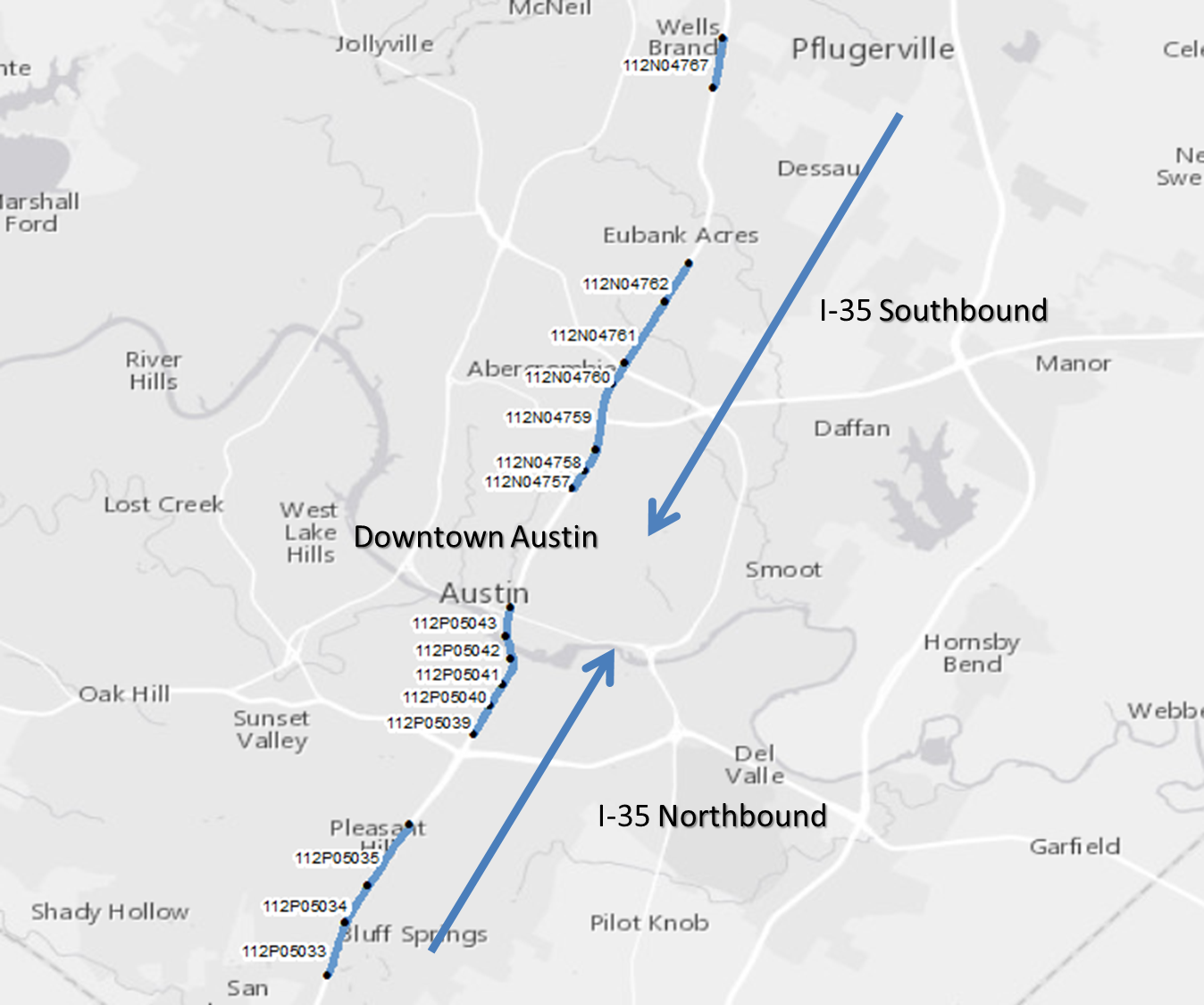}
\caption{The 15 road segments used in this research}
\label{location}
\end{figure}

The time-of-day travel time of TMC road segments vary quite substantially. \Cref{fig:2} plot time-of-day travel time for four typical TMCs, where two TMCs have short and long morning congestion periods but not in the afternoon, one TMC has afternoon congestion but not in the morning, and the other TMC has no clear congestion patterns. We extract all TMCs on the highway in the Austin Metropolitan Area that have congestion in the morning peak, in all 15 TMCs shown and labeled in \cref{location}.  The arrows show the direction of the traffic flow for those TMCs.  It is not surprising to see that all TMCs with traffic heading from suburb areas to downtown Austin are selected. For references we number the 15 road segments from 1 to 15, in its geographic order from north to south, i.e `112N04767' is numbered by 1 and `112P05033' is numbered by 15.

Next, we define congestion starting time (CST) and congestion period using the travel time data. For each TMC $i$ at a time $t$, suppose the travel time is $r_i^t$ and the TMC's free flow travel time is $FFTT_i \doteq \min_t\{r_i^t\}$.  The TMC is stationarily congested at time $t$ if,

%
%

\begin{enumerate}
\item $r_i^t/FFTT_i \geq 2$;
\item Condition 1 holds for at least 15 minutes (namely $r_i^{t+1}/FFTT_i \geq 2$, $r_i^{t+2}/FFTT_i \geq 2$).
\end{enumerate}

Condition 2 is added to ensure that the congestion is not transient nor random. Furthermore, we can extract congestion starting time and congestion duration directly from the travel time data.

\begin{figure}
\centering
\includegraphics[scale = 0.5]{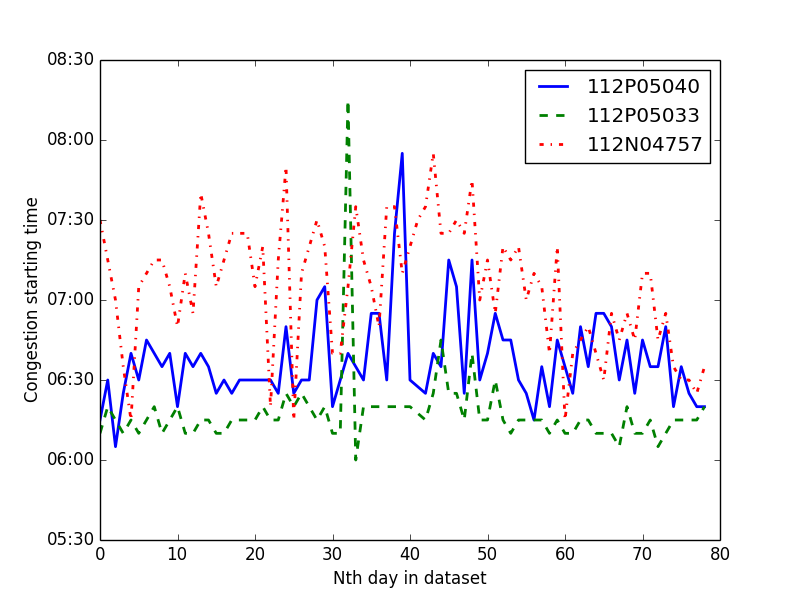}
\caption{Congestion starting time over 79 selected weekdays for three TMCs}
\label{fig:new1}
\end{figure}

\Cref{fig:new1} shows the extracted congestion starting time (CST) of three TMCs on all 79 weekdays. The TMC represented by the green line has a relatively low day-to-day variation of the congestion starting time in most of those days. However, there exist several days where the congestion started 30 minutes later than the typical time, or even two hours later on a particular day. For the TMCs presented by the blue line and red line, the congestion starting time can vary by more than one and a half hours from day to day. This research attempts to use electricity data to explain the day-to-day variation, and establish an accurate predictor at 6 am (or even earlier) of each day before the congestion starts.

\newpage

\section{Results and discussion}
\label{sec:5}
In this section we present and discuss prediction results. Depending on what results to present, we use different settings of training data set and testing data set.
\begin{itemize}
  \item All 79 weekdays are used to produce a predictor for interpreting correlations among features and the congestion. In this case, training and testing data are not separated.
  \item When assessing and comparing the overall performance of a predictor denoted by Root Mean Square Error (RMSE) and Mean Average Error (MAE), we use a two-level cross validation. All 79 weekdays are first divided into 3 folds (namely, 3-fold cross validation at the upper level). At each time, two of the three folds are picked out as the training data set, leaving the other fold as the testing data set. We apply 4-fold cross validation (namely the lower level cross validation) to the training data set to learn all parameters for the predictor (such as coefficients of features, and $\alpha$ of the LASSO model). Then the calibrated predictor is used to compute the RMSE or MAE on the testing data. The final RMSE or MAE are averaged from the upper level 3-fold cross validation.
  \item The two-level cross validation is best for providing the overall performance of a predictor with minimum overfitting. However, those parameters of the predictor can vary substantially by the training data set. Therefore, when demonstrating a particular predictor, we randomly divide the 79 weekdays into a training set of 60 weekdays and a testing set of 19 weekdays for each of the 15 TMCs. For each TMC, there may exist several weekdays that have no morning congestion. We filter out those weekdays when studying a predictor for this particular TMC. Thus, the number of weekdays used to train a model varies by TMC, but generally accounts for about 75\% of those weekdays. In this case, a predictor is learned based on the fixed training set, and tested using the fixed testing set.
\end{itemize}


\subsection{Predicting congestion starting time (CST) using aggregated features}

We first use those aggregated features, namely the portions of households in all ten electricity use patterns, to predict morning congestion starting time for each of the 15 TMCs. This subsection highlights the correlations between features and the congestion starting time, which uses all 79 weekdays in regressing the predictor.

\begin{figure}
\centering
\begin{subfigure}[b]{0.49\textwidth}
        \includegraphics[width=\textwidth]{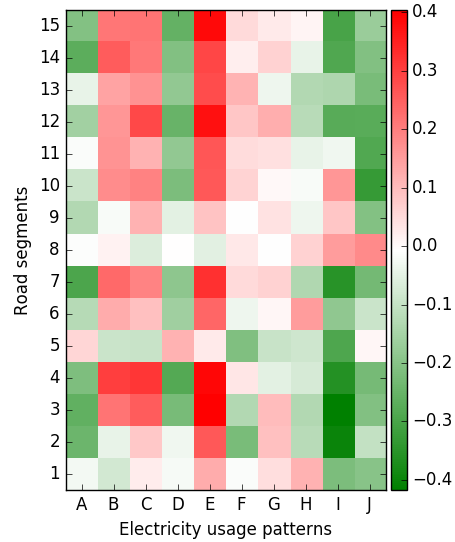}
        \caption{Coefficients}
\end{subfigure}
\begin{subfigure}[b]{0.47\textwidth}
        \includegraphics[width=\textwidth]{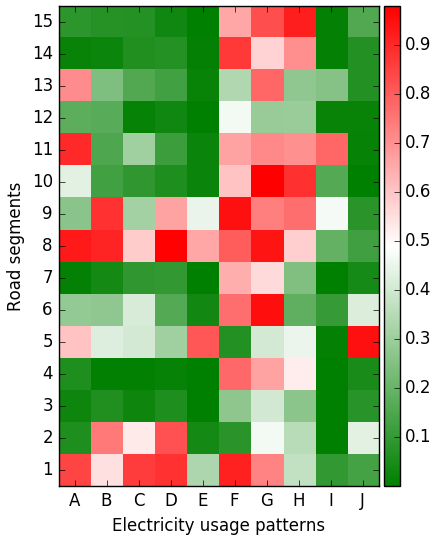}
                \caption{P-values}
\end{subfigure}
\caption{Regression results using aggregated features as a linear predictor}
\label{portion}
\end{figure}

\Cref{portion} shows the coefficients and their respective p-values for all 15 linear predictors. Patterns $B$, $C$, $E$, $F$ and $G$ have positive effects on the congestion starting time. Patterns $B$, $C$ and $E$ in \Cref{c:result} are households who steadily use electricity after midnight with an oscillating and declining usage over time. It is no surprise that more households in those patterns on a particular day lead to a later congestion starting time next time. We speculate that those households with those types of after-midnight activities are likely commuters. They are likely to commute later if under those patterns than if under patterns $A$, $I$ and $J$. Of those positively correlated patterns, $B$, $C$ and $E$ have higher positive effects than $F$ and $G$, and the effects on $F$ and $G$ are not statistically significant. $F$ and $G$ are households who use electricity intensively at around 1:00-1:30am, but the usage is very low at all other times. This type of midnight activities may not imply commuters, or not necessarily related to their travel activities.

Patterns $A$, $D$, $I$ and $J$ have negative effects on the congestion starting time, which are statistically significant for most TMCs. The effect by pattern $D$ is slightly milder. Patterns $A$, $I$ and $J$ each represents a group of households whose electricity use increases from 2 am and then declines before 6 am with possibly a particular work schedule in the early morning, consistent with the speculation that users get up in early morning and leave by 6 am for work. Pattern $D$ shows the electricity use is relatively high and generally stable from midnight to 6 am. However, it still implies a slight decline after midnight and a slight increase after 5 am. It is intuitive that more households under those four patterns imply an earlier departure time from home, thus possibly leading a earlier congestion starting time.

\begin{figure}[h]
            \begin{subfigure}[b]{0.33\textwidth}
        \includegraphics[width=\textwidth]{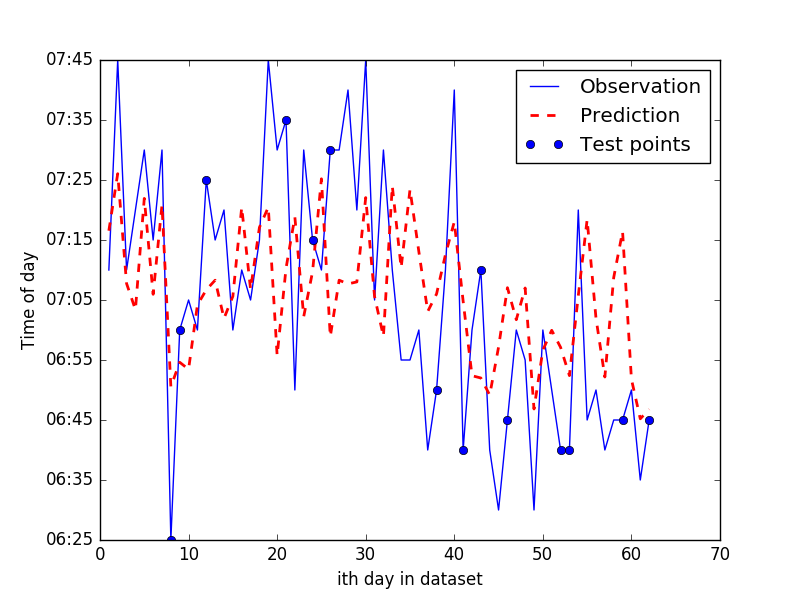}
        \caption{TMC 2, RMSE: 0.45 hr}
    \end{subfigure}
                \begin{subfigure}[b]{0.33\textwidth}
        \includegraphics[width=\textwidth]{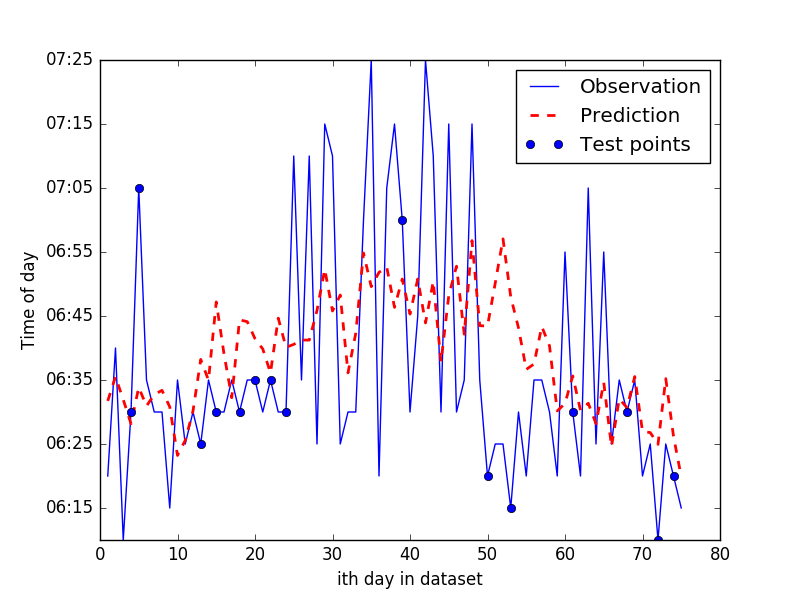}
        \caption{TMC 12, RMSE: 0.32 hr}
    \end{subfigure}
                    \begin{subfigure}[b]{0.33\textwidth}
        \includegraphics[width=\textwidth]{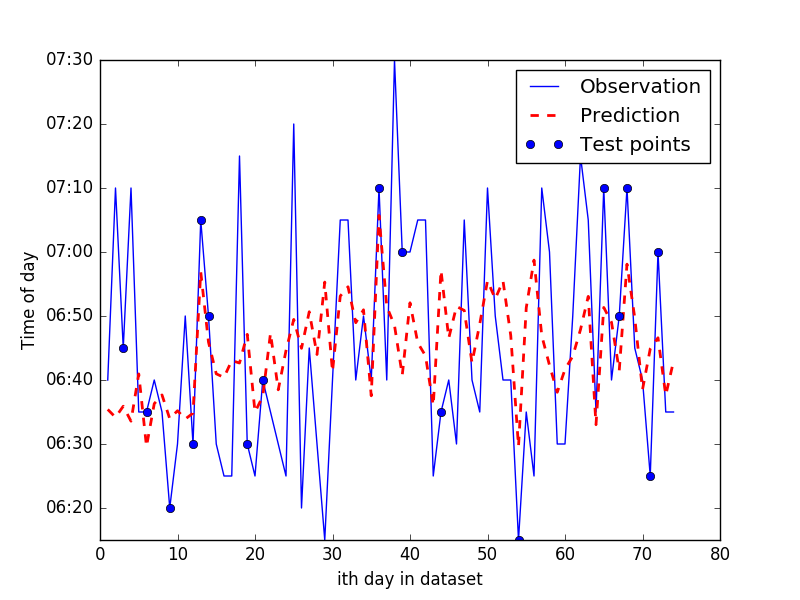}
        \caption{TMC 9, RMSE: 0.32 hr}
    \end{subfigure}
\caption{Prediction results using a linear regression model based on aggregated features}
\label{fig:portion}
\end{figure}

Though the effects of aggregated features discovered from the above linear regression model are promising, the prediction results are not perfect. The linear model learned from all 79 weekdays does not accurately predict the congestion starting time, even without leaving testing data out of training data. We divide data into training data set and testing data set, and regress the model on the training data set. The actual congestion starting time against the predict time, on weekdays of both training set and testing set, are plot in \Cref{fig:portion} for randomly chosen three TMCs.

The linear model captures overall trend of change in congestion starting time, but the predicted results can differ from the true value substantially on some days, especially for the testing data set. The maximum prediction error on a day can be as many as $50$ minutes.  In other words, aggregated features imply useful information regarding the commuters' activities, but the prediction may be further improved. Next we attempt to use disaggregate features that carry more household-level information than aggregate features, and traffic features.

\subsection{Predicting congestion starting time (CST) using disaggregate features: an example of overfitting}

Now we regress the congestion starting time by disaggregate features. The number of features are much greater than the sample size, $2898$ versus $79$. Thus, we use LASSO regularization described in Equation \ref{lasso} for each TMC, with a pre-selected shrinkage parameter ( i.e. $\alpha$ in Equation \ref{lasso}).

\begin{figure}[h]
\centering
            \begin{subfigure}[b]{0.45\textwidth}
        \includegraphics[width=\textwidth]{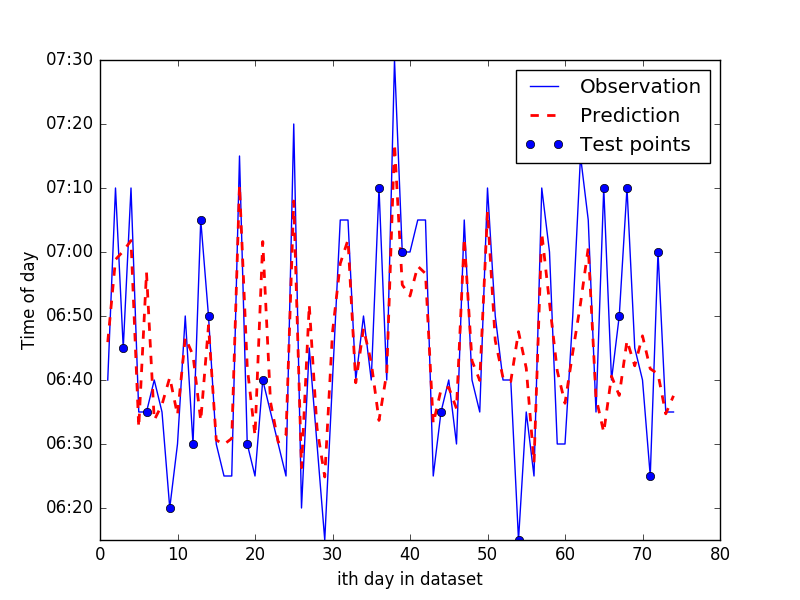}
        \caption{$\alpha = 0.01$, RMSE: 0.34 hr}
                \label{TMCb}
    \end{subfigure}
            \begin{subfigure}[b]{0.45\textwidth}
        \includegraphics[width=\textwidth]{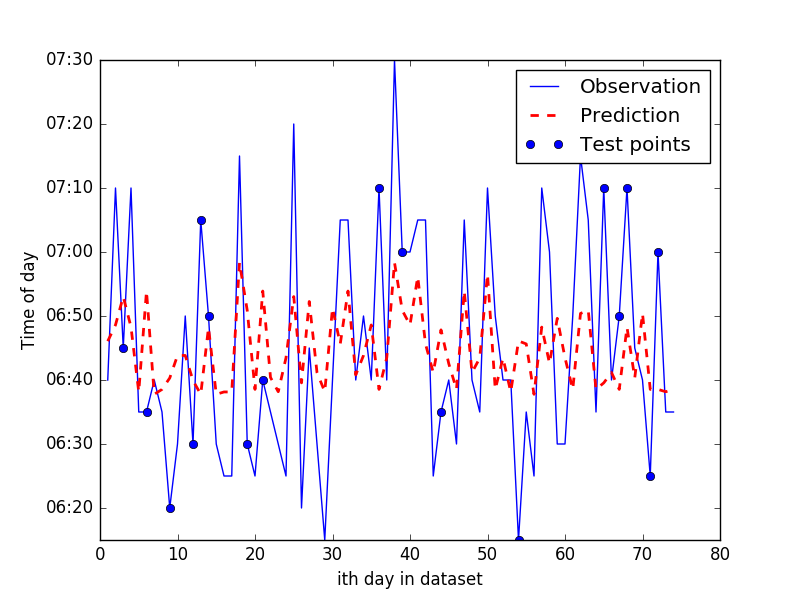}
        \caption{$\alpha = 0.033$, RMSE: 0.30 hr }
        \label{TMCc}
    \end{subfigure}
\caption{Example of TMC 9 `112P05042':  (a) and (b) show the raw data together with values predicted by the LASSO model with $\alpha=0.01$ and $0.033$ respectively.}
\label{example TMC}
\end{figure}

Using TMC 9 `112P05042' as an example, Figure  \ref{example TMC} shows predicted congestion starting time against actual ones, for both the training and testing weekdays. $\alpha=0.033$ implies a much more restrictive selection of variables than $\alpha=0.01$ where the latter tends to overfit the data. Comparing to Figure \ref{TMCc}, the predicted values in Figure \ref{TMCb} fit the data very well for those training days, but not for testing days. This is clearly attributed to overfitting. This issue is less pronounced in Figure \ref{TMCc}. On the other hand, the prediction result using disaggregate features in Figure \ref{TMCc} is better than the one using aggregate features in Figure \ref{fig:portion}. The maximum error, if using disaggregate features, is less than half an hour, and in most days the prediction error is less than $20$ minutes. This example implies that the importance of selecting a proper $\alpha$ to mitigate overfitting. It also shows that using disaggregated features has potential to improve the linear predictor.

In the next subsection, we use a two-level cross validation to develop models, and provide overall model performance.

\subsection{Comparing overall performance of predictors}

This subsection presents and compares the overall performance of predictors using aggregated features and disaggregated features of electricity use, as well as predictors that use only features derived from traffic data. This allows us to examine if electricity use data brings additional information to better predict congestion than using only real-time traffic data.

Two methods are used in this comparison to predict congestion using only real-time traffic data. First, we use the historical mean of congestion starting time (CST) measured in previous five weekdays to predict the CST on the next day. This is a naive method without requiring real-time traffic data access. It is referred to as ``historical mean'' later. The second method is to use the real-time travel time data to predict the change in travel time in near future, and then check if the predicted travel time reaches the two conditions discussed in Section \ref{sec:congest} to constitute a CST. We choose a commonly used time-series model, the ARMA model. A general ARMA model predicts the travel time at time $t$ by,
\begin{equation}
X_t = c  + \delta_t + \sum_{i=1}^p\phi_iX_{t-i} + \sum_{i=1}^q\theta_i\delta_{t-i}
\end{equation}
written as ARMA$(p,q)$, where $\delta_i$ are white noise terms for time interval $i$. $\phi_i,\theta_i$ are parameters for time $i$. $c$ is a constant. Studies have shown that ARMA model and its variants can make reasonable prediction of travel time in a short term such as 10 minutes \citep[e.g.,][]{billings2006application,guin2006travel}
However, in practice, 10 min is oftentimes too short to offer useful traveler information. A hours ahead prediction of travel time would be preferred. Just like the prediction using electricity use data before 6am, traffic data up to 6am will be used to form a time series to predict CST for a fair comparison. In other words, the same time-of-day  period of data (midnight to 6am), either electricity data or traffic data, will be used in this comparative study.

\begin{figure}
\centering
 \includegraphics[width=0.5\textwidth]{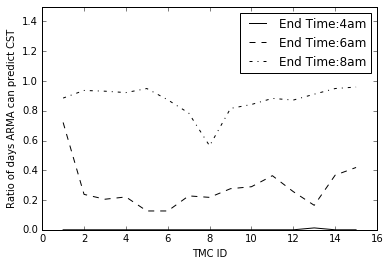}
 \caption{The portion of weekdays when ARMA can predict a morning congestion}
 \label{arma_endtime}
\end{figure}

Because the ARMA model predicts travel time which is then used to examine if congestion occurs and to infer the CST, it is possible that the ARMA model may predict no traffic congestion on a specific day. This is because congestion can grow very slowly prior to 6am on some weekdays, and ARMA is unable to pick up the traffic break-down using only the traffic data up to 6 am. \Cref{arma_endtime} shows the portion of weekdays when a CST can be retrieved from the ARMA model, if we use traffic data up to 4am, 6am and 8am, respectively. The order (namely $p$ and $q$) of ARMA is chosen through AIC (Akaike information criterion) criterion. Clearly ARMA would need traffic data up to 8am in order to predict a CST for most weekdays.  Recall in \Cref{fig:new1} that the morning congestion usually starts between 6 am and 8 am. Using travel time data up to 8 am to predict morning congestion is not useful at all. If we use data up to 4am, then for almost all weekdays, the ARMA cannot pick up a morning congestion. In this comparison, we compared ARMA with proposed models that use electricity-use data, all up to 6 am. The MSE and MAE for the ARMA model are computed for only those weekdays that ARMA can provide a CST. Therefore, the actual performance of ARMA should be considered much worse than those MSE and MAE presented below.

\begin{figure}[h]
            \begin{subfigure}[b]{0.49\textwidth}
                    \includegraphics[width=\textwidth]{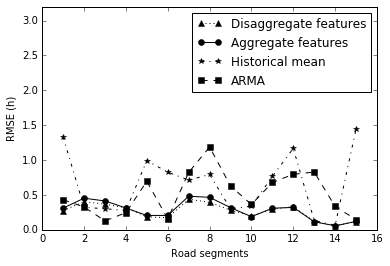}
        \caption{RMSE}
    \end{subfigure}
                \begin{subfigure}[b]{0.49\textwidth}
        \includegraphics[width=\textwidth]{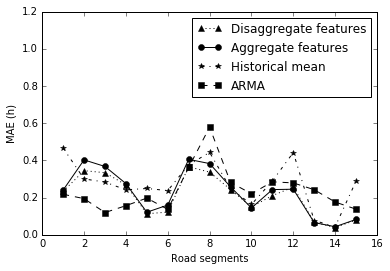}
        \caption{MAE}
    \end{subfigure}
\caption{Comparison among all four methods on predicting congestion starting time}
\label{comparison}
\end{figure}

\begin{table}
\centering
\caption{Predicting congestion starting time: comparison among different methods}
\label{tb:1}
\resizebox{\textwidth}{!}{%
\begin{tabular}{c|c|cc|cc|cc|cc|cc}
\hline
\multirow{2}{*}{TMC no.}& \multirow{2}{*}{ID}&\multicolumn{2}{ |c| }{Disaggregate features}& \multicolumn{2}{ |c| }{Aggregate features}&\multicolumn{2}{ |c| }{ARMA}&\multicolumn{2}{ |c| }{Historical mean}&\multicolumn{2}{|c}{Mixed features}\\
&&RMSE(h)&MAE(h)&RMSE(h)&MAE(h)&RMSE(h)&MAE(h)&RMSE(h)&MAE(h)&RMSE(h)&MAE(h)\\
\hline
112N04767 & 1 & 0.28 & 0.23 & 0.31 & 0.24 & 0.43 & 0.22 & 1.33 & 0.47 & 0.23 & 0.18 \\
112N04762 & 2 & 0.40 & 0.34 & 0.45 & 0.40 & 0.33 & 0.19 & 0.32 & 0.30 & 0.34 & 0.26 \\
112N04761 & 3 & 0.37 & 0.33 & 0.41 & 0.37 & 0.13 & 0.12 & 0.30 & 0.28 & 0.27 & 0.17 \\
112N04760 & 4 & 0.32 & 0.27 & 0.31 & 0.28 & 0.25 & 0.16 & 0.28 & 0.24 & 0.21 & 0.17 \\
112N04759 & 5 & 0.19 & 0.11 & 0.21 & 0.12 & 0.70 & 0.20 & 0.99 & 0.25 & 0.20 & 0.12 \\
112N04758 & 6 & 0.17 & 0.12 & 0.21 & 0.16 & 0.15 & 0.14 & 0.83 & 0.24 & 0.20 & 0.16 \\
112N04757 & 7 & 0.44 & 0.37 & 0.48 & 0.41 & 0.82 & 0.36 & 0.71 & 0.38 & 0.44 & 0.37 \\
112P05043 & 8 & 0.39 & 0.34 & 0.47 & 0.38 & 1.19 & 0.58 & 0.79 & 0.44 & 0.41 & 0.32 \\
112P05042 & 9 & 0.28 & 0.24 & 0.32 & 0.26 & 0.63 & 0.28 & 0.30 & 0.25 & 0.25 & 0.20 \\
112P05041 & 10 & 0.20 & 0.15 & 0.19 & 0.14 & 0.36 & 0.22 & 0.34 & 0.16 & 0.17 & 0.13 \\
112P05040 & 11 & 0.29 & 0.21 & 0.31 & 0.24 & 0.68 & 0.28 & 0.77 & 0.29 & 0.29 & 0.21 \\
112P05039 & 12 & 0.33 & 0.26 & 0.32 & 0.25 & 0.80 & 0.28 & 1.17 & 0.44 & 0.26 & 0.20 \\
112P05035 & 13 & 0.12 & 0.07 & 0.12 & 0.07 & 0.83 & 0.24 & 0.13 & 0.08 & 0.12 & 0.07 \\
112P05034 & 14 & 0.06 & 0.04 & 0.06 & 0.04 & 0.33 & 0.18 & 0.07 & 0.04 & 0.06 & 0.05 \\
112P05033 & 15 & 0.12 & 0.08 & 0.12 & 0.08 & 0.15 & 0.14 & 1.45 & 0.29 & 0.11 & 0.07 \\
\hline
\end{tabular}
}
\footnotesize{Note: ARMA is unable to pick up morning congestion in most days. The MSE and MAE for the ARMA model are computed for only those weekdays that ARMA provides a CST. Therefore, the actual performance of ARMA should be considered much worse than those MSE and MAE presented.}
\end{table}

%

\Cref{comparison} and \Cref{tb:1} show and contrast the RMSE and MAE of all four methods discussed above. Using disaggregate features of electricity-use data offers a more accurate prediction than using aggregate features for 9 out of 15 TMCs. This is no surprise since it carries more detailed information regarding usage patterns. In most TMCs, using electricity data is far more advantageous than using traffic data only. This result implies that electricity usage pattern is spatially and temporally correlated with highway usage, and it is possible to predict morning congestion from electricity use during midnight and early morning. The only exception is TMC 2 and TMC 3 where ARMA and historical mean provide better prediction. Note that in this case study, the locations of those households where electricity usage is acquired are unknown. Clearly the electricity use of those households does not bring in useful information to predict traffic in those two TMCs, comparing to use historical average. We speculate that either very few of those households have trips related to these two segments of highway during morning peak hours. It would be expected that location information regarding those households or data from additional households could potentially improve our model.


\subsection{Predicting congestion starting time (CST) using mixed features}
\begin{figure}[h]
            \begin{subfigure}[b]{0.49\textwidth}
                    \includegraphics[width=\textwidth]{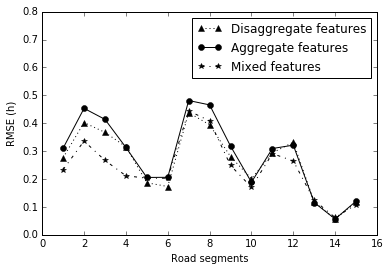}
        \caption{MSE}
    \end{subfigure}
                \begin{subfigure}[b]{0.49\textwidth}
        \includegraphics[width=\textwidth]{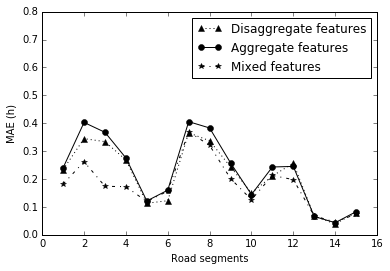}
        \caption{MAE}
    \end{subfigure}
\caption{Comparison three methods that use electricity data on predicting congestion starting time}
\label{fig:mix}
\end{figure}
Of all the four methods, using disaggregate features has the best performance overall. However, in practice, households can enter or quit the AMI program at any time, which can cause issues that the predictor using disaggregated features need to be re-tuned and updated at any of those occasions. On the other hand, a predictor using aggregate features will be less affected by households enrolling in AMI, since it only uses the portion of households in ten typical electricity use patterns. In addition, if the self-selection of households enrolling in AMI is consistent, we would expect the predictor using aggregated features to representative and perform well. \Cref{tb:1} shows that it is not much worse than using disaggregate features. Therefore, we would prefer to use aggregated features in practice to represent electricity use patterns, which, however, can be augmented with features extracted from traffic data.

We propose to use mixed features that combine both electricity usage information and traffic information. We first use real-time travel time up to a time, say 6am, to predict the CST with the ARMA model, denoted by $t^a$. By looking into historical data of CST, we obtain a time window $[t^+,t^-]$ that covers all CSTs historically.  If $t^a \in [t^+,t^-]$, then we simply append $t^a$ to those aggregate features for the predictor. If $t^a<t^+$ or $t^a > t^-$, we append $t^+$ and $t^-$, respectively, to those aggregate features. When ARMA does not provide a CST, it can be understood as the CST is infinitely late and $t^-$ is then used for the mixed features. 

The RMSE and MAE of the predictor using mixed features are also reported in \Cref{tb:1}. For a better illustration, they are also plot in \Cref{fig:mix} with results of predictors using aggregate features and disaggregate features. We can see that using mixed features outperforms using disaggregated features for most TMCs except TMC 6. The improvement on several TMCs is quite significant, such as TMC 3 and TMC 4. Overall, combining aggregated energy-use information and traffic information indeed provide the best performance. It is a very compelling results since the predictor can work well without utilizing personally identifiable information.   

\subsection{Predicting congestion duration}

\begin{figure}[h]
            \begin{subfigure}[b]{0.49\textwidth}
                    \includegraphics[width=\textwidth]{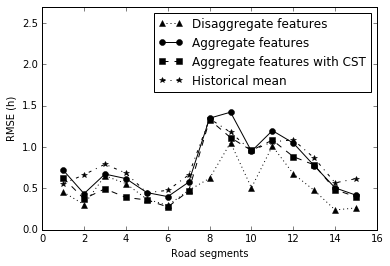}
        \caption{MSE}
    \end{subfigure}
                \begin{subfigure}[b]{0.49\textwidth}
        \includegraphics[width=\textwidth]{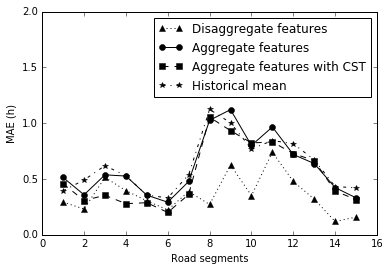}
        \caption{MAE}
    \end{subfigure}
\caption{Comparison among all four methods on predicting congestion duration}
\label{fig:duration}
\end{figure}

Now we predict the morning congestion duration using similar features. The predictors using aggregate features and disaggregate features, and historical means are used again. In addition, we create a new predictor that uses the predicted CST (from aggregated features) in addition to aggregate features. \Cref{fig:duration} plots the RMSE and MAE which are also reported in \Cref{tb:duration}.

Generally, the predictor with disaggregated features has reasonable results and considerably outperforms other predictors in all TMCs except TMC 3. While in most TMCs using aggregate features with and without CST information are very close to the predictor using the historical mean. It seems that neither the aggregate features nor the CST carries useful information to explain the morning congestion duration.

Comparing to predicting CST, predicting the duration has higher RMSE. Clearly the congestion duration is more challenging to predict, given the households information from midnight to 6am.  This is not surprising. The electricity use data from midnight to 6am can better explain the spatial and temporal distribution of travelers in the early morning than in the later morning. Congestion duration is likely to be affected by many factors other than the starting time of using highways, such as how long travel demand peaks and higher probability of incidents during morning congestion. Those add more complications to the prediction of duration than CST. 

\begin{table}
\centering
\caption{Predicting congestion duration: comparison among different methods}
\begin{tabular}{c|c|cc|cc|cc|cc}
\hline
\multirow{2}{*}{TMC no.}& \multirow{2}{*}{ID}&\multicolumn{2}{ |c| }{Disaggregate features}& \multicolumn{2}{ |c| }{Aggregate features}&\multicolumn{2}{ |c| }{Aggregate features + CST}&\multicolumn{2}{ |c }{Historical mean}\\
&&RMSE(h)&MAE(h)&RMSE(h)&MAE(h)&RMSE(h)&MAE(h)&RMSE(h)&MAE(h)\\
\hline
112N04767 & 1 & 0.46 & 0.29 & 0.72 & 0.51 & 0.63 & 0.46 & 0.56 & 0.39 \\
112N04762 & 2 & 0.31 & 0.23 & 0.44 & 0.36 & 0.38 & 0.31 & 0.67 & 0.49 \\
112N04761 & 3 & 0.65 & 0.51 & 0.67 & 0.54 & 0.50 & 0.35 & 0.79 & 0.62 \\
112N04760 & 4 & 0.56 & 0.40 & 0.62 & 0.53 & 0.39 & 0.28 & 0.69 & 0.53 \\
112N04759 & 5 & 0.37 & 0.30 & 0.45 & 0.36 & 0.36 & 0.29 & 0.45 & 0.36 \\
112N04758 & 6 & 0.30 & 0.22 & 0.40 & 0.29 & 0.28 & 0.20 & 0.48 & 0.33 \\
112N04757 & 7 & 0.48 & 0.39 & 0.58 & 0.48 & 0.47 & 0.36 & 0.66 & 0.54 \\
112P05043 & 8 & 0.62 & 0.27 & 1.35 & 1.03 & 1.33 & 1.05 & 1.34 & 1.12 \\
112P05042 & 9 & 1.05 & 0.63 & 1.42 & 1.12 & 1.12 & 0.93 & 1.18 & 1.00 \\
112P05041 & 10 & 0.51 & 0.35 & 0.95 & 0.80 & 0.96 & 0.82 & 0.96 & 0.77 \\
112P05040 & 11 & 1.01 & 0.74 & 1.20 & 0.97 & 1.09 & 0.83 & 1.06 & 0.84 \\
112P05039 & 12 & 0.68 & 0.48 & 1.05 & 0.72 & 0.89 & 0.73 & 1.09 & 0.82 \\
112P05035 & 13 & 0.48 & 0.32 & 0.78 & 0.64 & 0.78 & 0.67 & 0.87 & 0.67 \\
112P05034 & 14 & 0.24 & 0.12 & 0.51 & 0.42 & 0.49 & 0.39 & 0.57 & 0.43 \\
112P05033 & 15 & 0.26 & 0.16 & 0.42 & 0.33 & 0.40 & 0.32 & 0.62 & 0.42 \\
\hline
\end{tabular}
\label{tb:duration}
\end{table}

%

\subsection{Sensitivity analysis: how long of electricity use data do we need to predict CST?}
The prediction presented above is made at 6am, based on electricity data from midnight to 6 am. We would want to predict CST as early as possible. This subsection investigates how early in the morning we can make this CST prediction while preserving a reasonable prediction error. In other words, up to what time of electricity use data do we need in order to reliably predict CST. 

\begin{figure}[h]
\centering
\includegraphics[scale = 0.4]{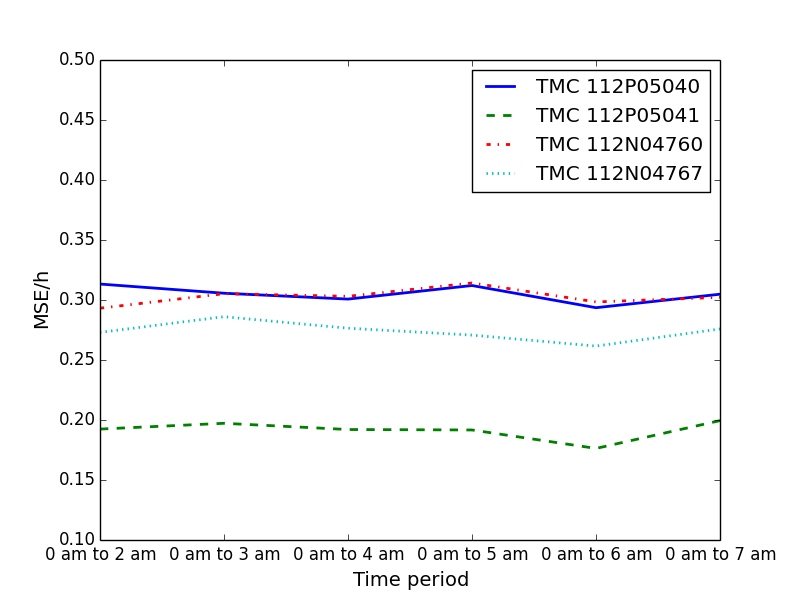}
\caption{Prediction performance with respect to the time periods of data used for prediction}
\label{period}
\end{figure}

\Cref{period} shows the overall performance of a predictor using disaggregated features with different choices of time periods on four representative TMCs. Generally the predictor performance is no so insensitive to the choice of time period. Using data from midnight to 6 am mildly outperforms other time periods. The prediction accuracy will slightly improve when using a longer period of data, but also declines when data beyond 6am is used. Note that the congestion starts some time between 6 am and 8 am, and oftentimes before 7 am. If the electricity usage data up to 7 am is used, it is possible that additional irrelevant information brought to the predictor. This is possibly due to increased noises of higher travel demand as time progresses in the morning. 

The change in performance with respect to time periods is relatively small. Even with the period from midnight to 2 am, the model can still make a reasonable prediction of CST, unlike models of ARMA or ``historical mean'' that uses traffic information where performance is heavily influenced by the choice of time periods. In fact, it is compelling that late night electricity use explains most of the information needed for understanding morning CST, so that the prediction can be made many hours ahead without deviating much from the true CST.



\subsection{Spatial prediction performance}
\begin{figure}[h]
\centering
\includegraphics[scale = 0.4]{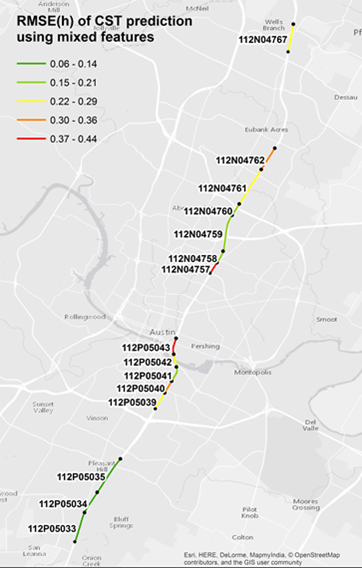}
\caption{The prediction RMSE of each TMC}
\label{rmse_tmc}
\end{figure}

This subsection examines the spatial variation of the predictor's performance. 
\Cref{rmse_tmc} shows the performance of all TMCs on a map, represented by the RMSE under those mixed features. The two TMCs closest to Austin downtown have the highest RMSE among all, implying our model may not work very well for highways within downtown. We would infer that the households in the AMI program are likely to reside in suburb area and use highways for commuting. It is also possible that the causes of congestion on roads near the downtown are likely to be complicated, which are not necessarily related to commuters' night indoor activities. 

In general the prediction on northbound TMCs is more accurate than that on southbound TMCs. We therefore speculate that more households in the AMI program live in the southern area than the northern area. While not using any spatially identifiable information from households, our models still produce reasonable results. We would also expect that the accuracy would improve if additional household location information can be incorporated into the prediction model.

\subsection{Critical households and electricity use patterns}
Last but not least, we investigate which households and what electricity use patterns are most critical in predicting CST. These are informed as a result of the Lasso regression. Specially we examine two things. First, are there any households that are consistently selected across all TMCs; and second, which electricity usage patterns are consistently selected across all TMCs. This provides insights for the value of information provided by all households/patterns in the AMI program. 

\begin{figure}[h]
            \begin{subfigure}[b]{0.49\textwidth}
                    \includegraphics[width=\textwidth]{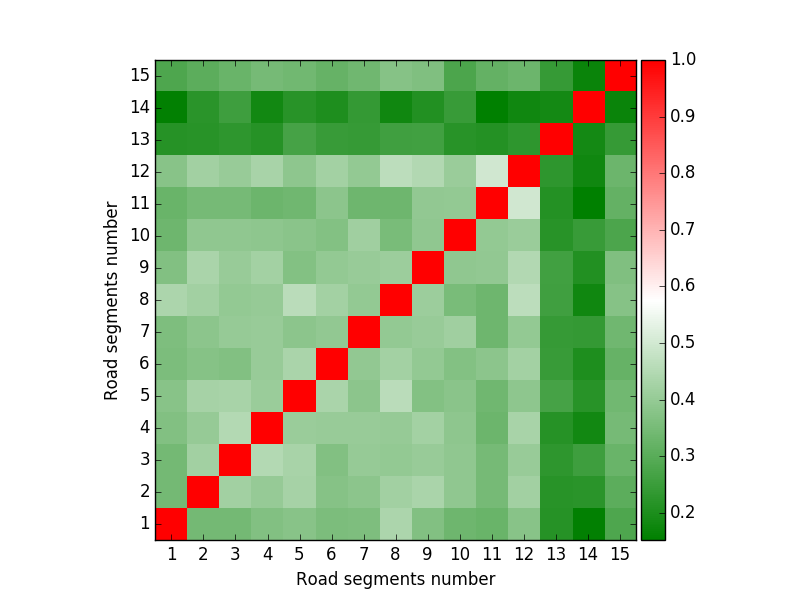}
        \caption{$\lambda$=0.0001}
    \end{subfigure}
                \begin{subfigure}[b]{0.49\textwidth}
        \includegraphics[width=\textwidth]{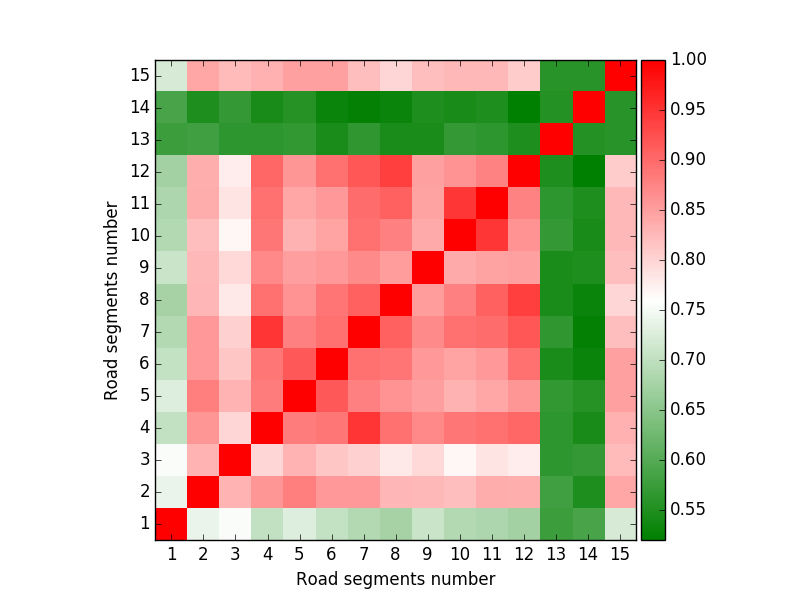}
        \caption{$\lambda$=0.0000001}
    \end{subfigure}
\caption{The Jaccard similarity coefficient of selected households for all TMC pairs}
\label{jsc}
\end{figure}

\Cref{jsc} computes the Jaccard similarity coefficient of all selected households for each TMC pairs, with respect to two possible values of $\lambda$ in Lasso. For two sets $\mathcal{A}$ and $\mathcal{B}$, the  Jaccard similarity coefficient is defined by,
\begin{align}
J(\mathcal{A},\mathcal{B}) = \frac{|\mathcal{A}\cap \mathcal{B}|}{|\mathcal{A}\cup \mathcal{B}| }
\end{align}

If the Jaccard coefficient of selected households for a pair of two TMCs is high, it implies that the congestion of these two TMCs are possibly due to the same group of households. As we can see from \cref{jsc}, in the square formed by TMCs 2 to 12, the coefficients are relatively high, while the coefficients are relatively low among TMCs 1,13,14 and 15. Those four TMCs (more than 10 miles away from downtown) are likely to be affected by a specific subset of households during the morning peak, while other TMCs that are closer to downtown are likely to be affected by a more broader group of households. Only a small fraction of travelers who use the outer TMCs contribute to the congestion of inner TMCs. Many of those travelers terminate their trips in the suburban areas, e.g. between TMC 12 and TMC 13. This is consistent with the fact that those highway segments do not typically have morning congestion, one of the reason why we cannot include them in this case study. 

\begin{figure}[h]
            \begin{subfigure}[b]{0.49\textwidth}
                    \includegraphics[width=\textwidth]{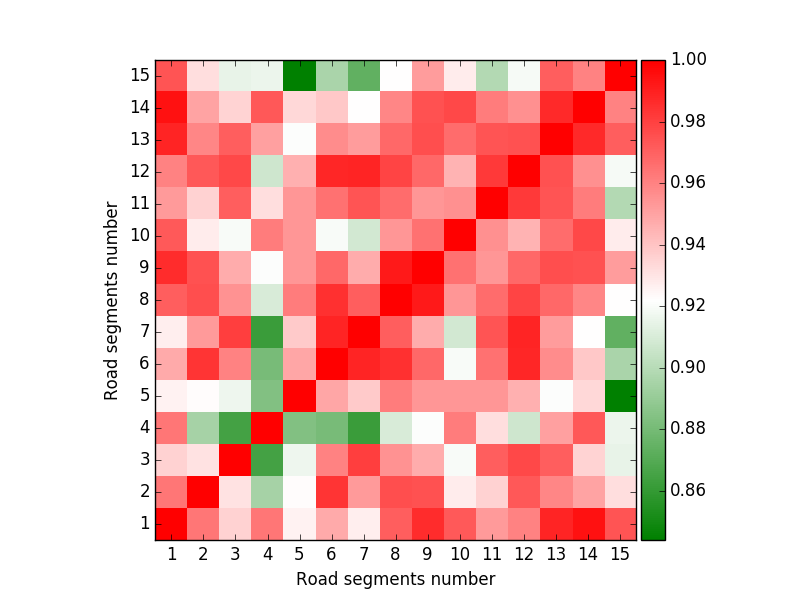}
        \caption{$\lambda$=0.0001}
    \end{subfigure}
                \begin{subfigure}[b]{0.49\textwidth}
        \includegraphics[width=\textwidth]{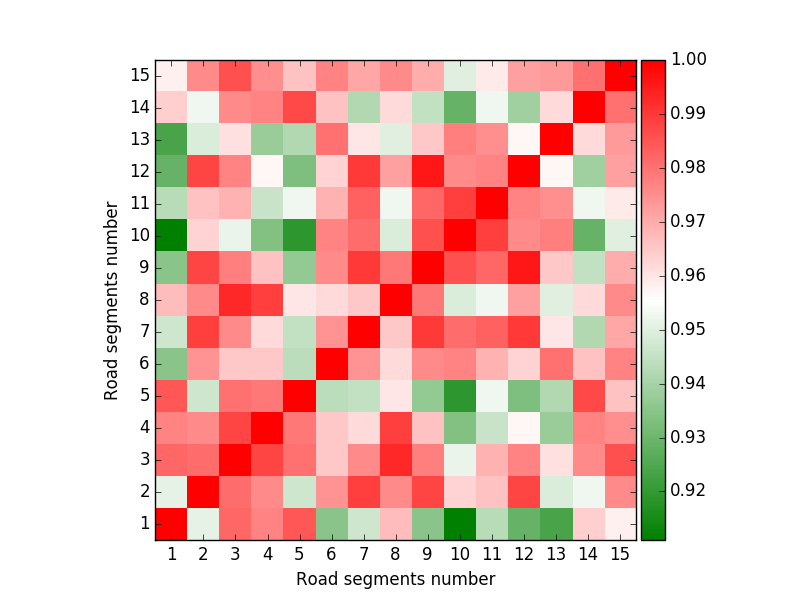}
        \caption{$\lambda$=0.00001}
    \end{subfigure}
\caption{The cosine similarity coefficient of selected electricity usage patterns for all TMC pairs}
\label{cos}
\end{figure}

We also plot the cosine similarity coefficient of the selected electricity usage patterns for all TMC pairs in \Cref{cos}. We count the number of households that fall in each of the ten patterns selected by LASSO to form a vector for each TMC, and use the cosine coefficient to measure their similarity. If the selected patterns between a pair of two TMCs are similar, than it implies that any selected electricity use patterns will have the same affect on morning congestion for the two TMCs. We can see that the similarity coefficient tends to be large (greater than 0.9) for all TMC pairs. This is different from the case of selected households. Congestion is likely to be related to a diversified group of households, but the effect of each electricity use pattern on all highway segments is very similar. Households can switch among electricity use patterns from day to day, and those patterns are essentially related to morning congestion. This makes sense that the effects of electricity usage patterns are spatially dependent. If an electricity usage pattern reveals that a person leaves home earlier, then all highway segments that this person uses will be affected in the same way. This is further verifies the importance of clustering of electricity use patterns.

\section{Conclusions}
\label{sec:6}

This paper explores the spatial and temporal correlation among usage patterns of energy systems and roadway systems. We propose a methodology along with data analytics for the City of Austin to predict morning congestion starting time and duration using the time-of-day electricity use data from 322 anonymous households with no spatial location information. Conceptually, the prediction model first categorizes users by their time-of-day electricity use patterns, learns the most critical features from those user categories that are related to traffic congestion, and finally builds a predictor dependent on electricity use features with parameters calibrated by real
data. The overall performance of the models is generated through Lasso selection and a two-level cross validation, and compared to the models using real-time traffic data only.

The results are very compelling and encouraging. We show that using sampled household-level electricity data from midnight to early morning, even from midnight to 2am, can reliably predict congestion starting time (CST) of many highway segments that are otherwise hard to predict using only real-time travel time data (through time series or the historical mean). We finally select an optimal predictor that uses electricity use information at the aggregated level, augmented with real-time traffic information to make the prediction of morning congestion start time at 6am. The predictor does not require knowing any personally identifiable information from households.

We found that 8 out of the 10 electricity use patterns (by the data from midnight to 6am) have significant affects on morning congestion on highways. Some are negative effects, represented by an early spike of electricity use followed by a drastic drop that could imply departure from home. Others are positive effects, represented by a late night spike of electricity use implying possible late night activities that possibly lead to late departure from home. The results also imply that morning congestion is likely to be related to a diversified group of households, but the effect of each electricity use pattern on all highway segments is very similar. Households can switch among electricity use patterns from day to day, and those patterns are essentially related to morning congestion. This makes sense that the effects of electricity usage patterns are spatially dependent. If an electricity usage pattern reveals that a person leaves home earlier, then all highway segments that this person uses will be affected in the same way.

We also found that late night electricity usage can explain most of the information needed for understanding morning CST, so that the prediction can be made many hours ahead, such as 2am in this case study,  without deviating much from the true CST. Using data from midnight to 6 am mildly outperforms using other time periods. The prediction accuracy will slightly improve when using a longer period of data, but also declines when data beyond 6am is used. Note that the congestion starts some time between 6 am and 8 am, and oftentimes before 7 am. If the electricity usage data up to 7 am is used, it is possible that additional irrelevant information brought to the predictor. This is possibly due to increased noises of higher travel demand as time progresses in the morning.

There are several limitations of this research. Limited by the accessibility to data set and seasonable effects,  the number of data used in the training the predictor is not large, especially comparing to the high dimension of the features. Including more data may help better train the predictor. Other factors together with the electricity usage data, e.g., the weather and incidents data that are known to effectively impact the traffic, could also potentially improve the performance. In this case study, using electricity use data to predict congestion works very well on some highway segments, but not so great on others. This may be due to the sampling bias of households enrolled in the AMI program, small sample size for the City of Austin, and unknown locations of those households.  We would expect that the accuracy would improve if additional household location information can be incorporated into the prediction model. In addition, this case study only works for the City of Austin since it is extremely difficulty to obtain electricity use data. We plan to actively acquire additional electricity use data from other cities to further validate our models.



\cleardoublepage

\bibliography{report}
\cleardoublepage
\appendix
\cleardoublepage

\end{document}